\documentclass[12pt]{article}

\usepackage[utf8]{inputenc}
\usepackage[T1]{fontenc}
\usepackage{authblk} 
\usepackage{amsmath,amsfonts,amssymb,amsthm}
\usepackage{dsfont} 
\usepackage{color}
\usepackage[dvipsnames]{xcolor} 
\colorlet{mygreen}{green!60!black}
\colorlet{mygrey}{white!60!black}
\usepackage{graphicx} 
\usepackage{float}  
\usepackage{booktabs} 
\usepackage{setspace} 
\usepackage{multicol} 
\usepackage{url} 
\usepackage{hyperref}

\usepackage{wrapfig}
\usepackage{lscape}
\usepackage{rotating}
\usepackage{epstopdf}

\usepackage{tikz}
\usetikzlibrary{shapes,positioning} 
\usetikzlibrary{calc} 
\usetikzlibrary {arrows.meta} 

\newlength{\wirespace} 
\newlength{\wirespaceh} 
\newlength{\wirespaceht} 
\newlength{\wirespacehb} 
\newlength{\boxsize}
\newlength{\messpace} 
\usepackage{geometry} 
\usepackage[font=small,labelfont=bf]{caption} 
\usepackage{palatino} 

\numberwithin{equation}{section} 

\theoremstyle{definition}

\theoremstyle{theorem}

\def \be {\begin{equation}} 
\def \ee {\end{equation}}
\def \bes {\begin{equation*}}
\def \ees {\end{equation*}}
\def \baa {\begin{align}}
\def \eaa {\end{align}}
\def \baas {\begin{align*}}
\def \eaas {\end{align*}}
\def \bea {\begin{eqnarray}}
\def \eea {\end{eqnarray}}
\def \beas {\begin{eqnarray*}}
\def \eeas {\end{eqnarray*}}

\newcommand{\nib}[1]{\noindent\textbf{#1}}

\newcommand{\ct}[1]{\begin{center}\textit{#1}\end{center}}

\newcommand{\ket}[1]{\rvert#1\rangle}
\newcommand{\bra}[1]{\langle #1\rvert}

\newcommand{\tr}{\mathrm{tr}_}

\newcommand{\ketbra}[2]{| #1 \rangle \langle #2 |}
\newcommand{\comp}[1]{\overline{#1}}
\newcommand{\cnot}{\text{Cnot}}

\newcommand{\one}{\mathds{1}}

\newcommand{\bol}{\boldsymbol} 

\newcommand{\Dxq}{\mathcal D_x^{\vec \varphi} (\bol q_1)} 
\newcommand{\Dzq}{\mathcal D_z^{\vec \varphi} (\bol q_1)}
\newcommand{\Uf}{\mathsf U} 
\newcommand{\calO}{\mathcal{O}}

\title{Teleportation Revealed}
\author{Charles Alexandre B\'edard}
\affil{\small {Universit\`a della Svizzera italiana}\\
\footnotesize \emph{charles.alexandre.bedard@usi.ch}}
\date{April 2023\vspace{-10pt}}

\begin{document}
\maketitle
\thispagestyle{empty} 

\abstract{\noindent 
Quantum teleportation is the name of a problem:
%
How can the 
 real-valued parameters encoding the state at Alice’s location make their way to Bob’s location via shared entanglement and only two bits of classical communication?
Without an explanation, teleportation appears to be a conjuring trick.
Investigating the phenomenon with Schrödinger states and reduced density matrices shall always leave loose ends because they are not local and complete descriptions of quantum systems.
Upon demonstrating that the Heisenberg picture admits a local and complete description, Deutsch and Hayden rendered its explanatory power manifest by revealing the trick behind teleportation, namely, by providing an entirely local account.
Their analysis is re-exposed and further developed.

\section{Introduction}

In the context of the recent Nobel Prize of physics, the \emph{Scientific American} published~\cite{notloacllyreal} an article titled ``The Universe Is Not Locally Real, and the Physics Nobel Prize Winners Proved It''.
I could dedicate my piece to the fact that we do not ``prove'' such claims in science or to the fact that the universe is real.
Instead, I will address the question of locality, which creates tremendous confusion in the community of quantum foundations.
The key message that I want to advocate is that
\emph{quantum systems can be described in a local and complete way, and we should do so.} 

Since Bell~\cite{bell1964} the term locality, more often seen negated, has come to mean compatibility with an underlying explanation by local hidden variables.
However, local hidden variables are only one way in which locality can be instantiated, whose full generality is captured by \emph{Einstein's locality}~\cite{schilppalbert1970}: ``the real factual situation of system~$A$ is independent of what is done with the system~$B$, which is spatially separated from the former.'' 
Scientific theories are tentative descriptions of the real factual situation, thus,
Einstein's locality can be lifted (and slightly generalized) into a criterion for theories:
\emph{The description of system~$A$ is independent of what is done with system~$B$, which is dynamically isolated from the former.} 

If Alice and Bob share an entangled pair of particles in a pure state, the reduced density matrices provide a local mode of description. 
Indeed, an action by Bob on his quantum system shall alter its density matrix, but Alice's remains unchanged. 
However, reduced density matrices are not a \emph{complete} mode of description, a sufficient definition of which is that 
\emph{the distribution of any joint measurement can be computed from the individual descriptions}.
The object that encompasses the distribution of any joint measurement is the global state vector,
which cannot be retrieved from the reduced density matrices since too much information has been traced out. 

The global state vector can also serve as a mode of description in which both Alice and Bob take it as the description of their own system.
It is complete; however, it is not local, for
if Bob makes a local change to his quantum system, it alters its description, which is fine, but it also alters the description of Alice's system.
We seem to be stuck in a dichotomy: quantum systems are described either locally or completely, and the appropriate description is chosen based on the problem at hand.

However, this dichotomy was proven false in 2000 by Deutsch and Hayden~\cite{deutsch2000information}, who showed that the Heisenberg picture admits \emph{descriptors}, which fulfil both Einstein's locality and completeness.
To insist, Bob's action on his system alters its descriptor but leaves the descriptor of Alice's system invariant; 
moreover, the global state vector can be recovered from the pair of descriptors.
Note that the existence of such a local mode of description in quantum theory makes the theory local:
the existence of non-local ways in which the theory can be expressed is irrelevant\footnote{
Terminology-wise, that the wave function is not a local and complete description has been referred to as its \emph{nonseparability}~\cite{healey1991holism, 
 wallace2010quantum}. 
Descriptors are a separable account, and therefore, quantum theory is separable.}.

In the more recent literature, there are two other approaches to what can be called quantum locality: the proposals of Raymond-Robichaud. 
\emph{Evolution matrices}~\cite{raymond2021local} are framed within the quantum formalism. 
\emph{Noumenal states}~\cite{raymond2017equivalence}, on the other hand, take a more general approach, as they apply to a class of theories for which operations form a group. 
Since quantum theory qualifies, noumenal states can be instantiated in quantum theory.
I proved~\cite{bedard2021cost} all these modes of description to be formally equivalent, and investigated some of the consequences of embracing these modes of description as an account of reality. 

We have been inundated with numerous alternative approaches to quantum theory, and it may appear that I am advocating yet another. 
But that is not what it is.
The proposal by Deutsch and Hayden hinges on previous work by Gottesman on the Heisenberg representation of quantum computers~\cite{gottesman1998heisenberg}, i.e. the Heisenberg picture of quantum theory applied to networks of qubits.
This picture is the way in which the theory was discovered~\cite{heisenberg1925quantum} in 1925, with the usual dynamical variables being promoted to dynamical operators. 
Consequently, what I shall present here is not about interpretation; it is about the mathematical formalism of quantum theory. 
It is the Heisenberg picture of unitary quantum theory.

If one finds locality and completeness not good enough reasons to adopt Heisenberg-picture descriptors as our tentative best account of quantum systems, then how they can be put to work might be more persuasive.
To demonstrate the explanatory power of Heisenberg-picture descriptors, I delve into a problem; the problem of teleportation~\cite{bennett1993teleporting}. 
\emph{How can the real-valued parameters encoding the state at Alice’s location make their way to Bob’s location via shared entanglement and only two bits of classical communication?} 

In this piece, 
I re-expose and further develop Deutsch and Hayden's solution.
If teleportation felt like a magic trick, they unveiled it.
After an overview of teleportation in the Schrödinger picture~(\S\ref{sec:tele}) and of Heisenberg-picture descriptors~(\S\ref{sec:heis}), teleportation is revisited in the light of descriptors~(\S\ref{sec:heistele}).
Then, it is argued that more than two real-valued parameters are ``teleported,'' as descriptors also encompass counterfactual descriptive elements (\S\ref{sec:morethan}). A conclusion (\S\ref{sec:conclusion}) and a discussion (\S\ref{sec:discussion}) follow.  

\section{The Usual Take on Teleportation}\label{sec:tele}

The teleportation protocol --- whose textbook 
appearance is displayed in Figure~\ref{fig:convtele} --- starts by preparing a pair of entangled qubits in the $\ket{\Phi^+}$ state\footnote{The four Bell states are $
\ket{\Phi^{\pm}} = \frac{\ket{00} \pm \ket{11} }{\sqrt 2}$ and
$\ket{\Psi^{\pm}} = \frac{\ket{01} \pm \ket{10} }{\sqrt 2}$.}, which is then shared between Alice and Bob. 
Then, or in the meantime, Alice prepares a qubit in the state to be teleported,~$\ket \psi = \alpha \ket 0 + \beta \ket 1$.
Alice might put her personal diary into~$\alpha$ and~$\beta$, 
as there is plenty of room in~$\alpha$ and~$\beta$;
being complex numbers, they can encode infinitely many bits. 
Since all qubits involved are initiated in their~$\ket 0$ state, Alice's preparation is seen as an operation~$U$, which takes~$\ket 0$ to~$\ket \psi$.
She then performs a Bell measurement between her shared entangled state and her prepared qubit.
It yields two classical bits of output that she communicates to Bob over a classical communication channel, like a telephone.
Bob then manipulates his 
qubit in accordance with the two bits that he receives.
He will, or not, apply the~$X$ gate; and he will, or not, apply the~$Z$ gate.
After Bob's processing of his system, its corresponding state is~$\alpha \ket 0 + \beta \ket 1$.
The very fact that a phenomenon is called ``teleportation'' underlines its puzzlement:
\ct{How do $\alpha$ and $\beta$ make their way from Alice to Bob?}
Let me insist that~$\alpha$ and~$\beta$, in principle, contain infinitely many bits, but only two classical bits are communicated.

\tikzset{meter/.append style={draw, inner sep=0, rectangle, font=\vphantom{A}, minimum width=1.25\boxsize, minimum height=\boxsize, 
 path picture={\draw[black] ([shift={(.08,.24)}]path picture bounding box.south west) to[bend left=50] ([shift={(-.08,.24)}]path picture bounding box.south east);\draw[black,-latex] ([shift={(0,0.15)}]path picture bounding box.south) -- ([shift={(.3,-0.1)}]path picture bounding box.north);}}}
%

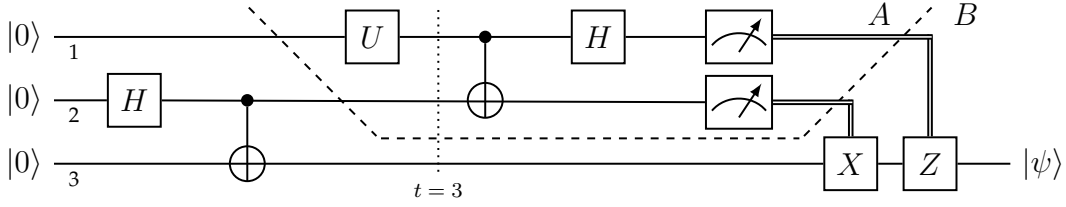
\begin{figure}[h]
	\centering
	\begin{tikzpicture} [scale=1, line width=0.75]
	
		\node (debut3) at (0,0){};
		\node (debut2) at (0, \wirespace){};
		\node (debut1) at (0, 2\wirespace){};
		
		\node [left = of debut3] (ket3) {$\ket 0$};
		\node [left = of debut2] (ket2) {$\ket 0$};
		\node [left = of debut1](ket1) {$\ket 0$};

		\node[below right = -6pt and -36pt of debut1] (w1) {\scriptsize{1}};
		\node[below right = -6pt and -36pt of debut2] (w2) {\scriptsize{2}};
		\node[below right = -6pt and -36pt of debut3] (w3) {\scriptsize{3}};
		
		\node[draw, rectangle, minimum width=\boxsize, minimum height=\boxsize, right=20pt of ket2] (H1){$H$};
		
		\node[draw, circle, fill=black, minimum width=0.14cm, inner sep=0pt, right= 30pt of H1] (C1) {}; 
		\draw let  
    			\p1 = (C1) in
      			(\x1,0) circle (0.33\boxsize)
			node (T1){}; 
		\draw let 
			\p1 = (T1) in
			(C1) -- (\x1, \y1-0.33\boxsize); 
		
		\node[draw,rectangle, minimum width=\boxsize, minimum height=\boxsize, 		right=(90pt+\boxsize) of ket1] (U) {$U$};
		
		\node[draw, circle, fill=black, minimum width=0.14cm, inner sep=0pt, right= 30pt of U] (C2) {}; 
		\draw let  
    			\p1 = (C2) in
      			(\x1,\wirespace) circle (0.33\boxsize)
			node[inner sep = 0] (T2){}; 
		\draw let 
			\p1 = (T2) in
			(C2) -- (\x1, \y1-0.33\boxsize); 
			  
		\node[draw,rectangle, minimum width=\boxsize, minimum height=\boxsize, right = 30pt of C2] (Hbell){$H$};
		
		 \node[meter , right=30pt of Hbell] (mes1) {};
		\node[meter , below=(\wirespace-\boxsize) of mes1] (mes2) {};
		
		\node[right=30pt of mes2, inner sep = 0] (coinx) {};
		\node[right=60pt of mes1, inner sep = 0] (coinz) {};
		
		\node[draw, rectangle, minimum width=\boxsize, minimum height=\boxsize, inner sep = 0, right=291pt of ket3] (X){$X$};
		\draw[double] 
			(X) -- ++(0,\wirespace+0.3pt);
			
		\node[draw, rectangle, minimum width=\boxsize, minimum height=\boxsize, right=321pt of ket3] (Z){$Z$};
		\draw[double] 
			(Z) -- ++(0,2\wirespace+1pt);
		
		\node[right = 20pt of Z] (fin3) {$\ket \psi$};
		
		\node (marker) at ([xshift=25pt, yshift=15pt]U){}; 
		\node (t3) at ([xshift=25pt, yshift=-2\wirespace-10pt]U) {\scriptsize{$t=3$}};
		\draw[dotted] (t3) -- (marker);
		
		  \node[inner sep = 0] at (40pt, 2.5\wirespace) (p1){};
		  \node[inner sep = 0] at (90pt, 0.4\wirespace) (p2){};
		  \node[inner sep = 0] at (250pt, 0.4 \wirespace)(p3){};
		  \node[inner sep = 0] at (300pt, 2.5\wirespace) (p4){};
		  
		  \node[below left = -2pt and 15pt of p4, inner sep = 0] (A) {$A$};
		  \node[below right = -2pt and 6pt of p4, inner sep = 0] (B) {$B$};

		 \draw[dashed] (p1) -- (p2); 
		 \draw[dashed] (p2) -- (p3); 
		 \draw[dashed] (p3) -- (p4); 
		
		\draw (ket1) -- (U);
		\draw (U) -- (Hbell);
		\draw (Hbell) -- (mes1);
		\draw[double] (mes1) -- (coinz);
		\draw (ket2) -- (H1);
		\draw (H1) -- (mes2);
		\draw[double] (mes2) -- (coinx);
		\draw (ket3) -- (X);
		\draw (X) -- (Z);
		\draw (Z) -- (fin3);
	\end{tikzpicture}
	\caption{Teleportation in a yet-to-be-defined quantum--classical dualistic theory.}
	\label{fig:convtele}
\end{figure}

To review the computation in the Schrödinger picture,
it is expositive to re-express the global state after $\ket{\Phi^+}$ 
 and~$\ket \psi$ have been prepared.
Calling this time $t=3$ and disregarding normalization,
\beas 
\ket{\Psi_3}&=&
\rvert \psi \rangle \otimes \rvert \Phi^+ \rangle\\
&=&  \left( \alpha \rvert 0 \rangle + \beta \rvert 1 \rangle \right) \otimes \left( \rvert 00 \rangle  + \rvert 11 \rangle \right )\\
&=& \alpha \rvert 000 \rangle  + \beta \rvert 100 \rangle  + \alpha \rvert 011 \rangle  + \beta \rvert 111 \rangle \\
&=& \alpha \left(\rvert \Phi^+ \rangle + \rvert \Phi^- \rangle\right)  \rvert  0 \rangle  + \beta \left(\rvert \Psi^+ \rangle - \rvert \Psi^- \rangle\right)  \rvert0 \rangle  + \alpha \left(\rvert \Psi^+ \rangle + \rvert \Psi^- \rangle\right) \rvert1 \rangle  + \beta \left(\rvert \Phi^+ \rangle - \rvert \Phi^- \rangle\right)  \rvert1 \rangle \\
&=& \rvert \Phi^+ \rangle 
\left( \alpha  \rvert  0 \rangle + \beta \rvert 1 \rangle \right) +
\rvert \Phi^- \rangle 
\left( \alpha \rvert  0 \rangle  - \beta \rvert 1 \rangle \right)
+
\rvert \Psi^+ \rangle 
\left( \beta \rvert 0 \rangle + \alpha \rvert 1 \rangle \right)
+
\rvert \Psi^- \rangle  
\left( -\beta \rvert0 \rangle + \alpha \rvert 1 \rangle  \right)\\
&=& \rvert \Phi^+ \rangle 
\rvert  \psi \rangle
+
\rvert \Phi^- \rangle 
Z \rvert  \psi \rangle
+
\rvert \Psi^+ \rangle 
X \rvert  \psi \rangle
+
\rvert \Psi^- \rangle  
XZ\rvert  \psi \rangle \,.
\eeas
Expressing $\ket{\Psi_3}$ in such a way helps verifying the rest of the protocol: 
the Bell measurement distinguishes the four possible terms and provides the information about which correction needs to be done on Bob's system to recover~$\ket\psi$.
Verifying that the protocol achieves the purported functionality does not amount to explaining how it works.
If the flexibility 
 in the way~$\ket{\Psi_3}$ can be expressed is useful for the verification of the protocol, it is disastrous to provide an explanation of the transmission of~$\alpha$ and~$\beta$, for the mathematical equality between an expression describing~$\alpha$ and~$\beta$ at Alice's location and an expression displaying them to be at Bob's location annihilates the hopes to locate the information accurately in the state vector.

Importantly, embracing unitary quantum theory does not lead to much progress.
In fact, records of the measurement outcome 
can be appended as displayed in Figure~\ref{fig:unitary}. 
In this quantum setting, the state vector evolves according to 
\bea
\ket{\Psi_3} &=&
\ket{00} \left( \rvert \Phi^+ \rangle 
\rvert  \psi \rangle
+
\rvert \Phi^- \rangle 
Z \rvert  \psi \rangle
+
\rvert \Psi^+ \rangle 
X \rvert  \psi \rangle
+
\rvert \Psi^- \rangle  
ZX\rvert  \psi \rangle \right)\nonumber \\
\ket{\Psi_5}&=& \ket{00} \left( \rvert 00 \rangle 
\rvert  \psi \rangle
+
\rvert 10 \rangle 
Z \rvert  \psi \rangle
+
\rvert 01 \rangle 
X \rvert  \psi \rangle
+
\rvert 11 \rangle  
ZX\rvert  \psi \rangle \right)\nonumber \\
\ket{\Psi_7}&=& \ket{00}\rvert 00 \rangle 
\rvert  \psi \rangle
+
\ket{10}\rvert 10 \rangle 
Z \rvert  \psi \rangle
+
\ket{01}\rvert 01 \rangle 
X \rvert  \psi \rangle
+
\ket{11}\rvert 11 \rangle  
ZX\rvert  \psi \rangle\nonumber \\
\ket{\Psi_9}&=& \left( \ket{00}\rvert 00 \rangle 
+
\ket{10}\rvert 10 \rangle 
+
\ket{01}\rvert 01 \rangle 
+
\ket{11}\rvert 11 \rangle  \right) \rvert  \psi \rangle  \,. \label{eq:univ}
\eea
\begin{figure}[h]
	\centering
	\begin{tikzpicture} [scale=1, line width=0.75]
	
		\node (debut3) at (0,0){};
		\node (debut2) at (0, \wirespace){};
		\node (debut1) at (0, 2\wirespace){};
		
		\node [left = of debut3] (ket3) {$\ket 0$};
		\node [left = of debut2] (ket2) {$\ket 0$};
		\node [left = of debut1](ket1) {$\ket 0$};

		\node[below right = -6pt and -36pt of debut1] (w1) {\scriptsize{1}};
		\node[below right = -6pt and -36pt of debut2] (w2) {\scriptsize{2}};
		\node[below right = -6pt and -36pt of debut3] (w3) {\scriptsize{3}};
		
		\node[draw, rectangle, minimum width=\boxsize, minimum height=\boxsize, right=20pt of ket2] (H1){$H$};
		
		\node[draw, circle, fill=black, minimum width=0.14cm, inner sep=0pt, right= 30pt of H1] (C1) {}; 
		\draw let  
    			\p1 = (C1) in
      			(\x1,0) circle (0.33\boxsize)
			node (T1){}; 
		\draw let 
			\p1 = (T1) in
			(C1) -- (\x1, \y1-0.33\boxsize); 

		\node[draw,rectangle, minimum width=\boxsize, minimum height=\boxsize, 		right=(90pt+\boxsize) of ket1] (U) {$U$};
		
		\node[draw, circle, fill=black, minimum width=0.14cm, inner sep=0pt, right= 30pt of U] (C2) {}; 
		\draw let  
    			\p1 = (C2) in
      			(\x1,\wirespace) circle (0.33\boxsize)
			node[inner sep = 0] (T2){}; 
		\draw let 
			\p1 = (T2) in
			(C2) -- (\x1, \y1-0.33\boxsize); 
			  
		\node[draw,rectangle, minimum width=\boxsize, minimum height=\boxsize, right = 30pt of C2] (Hbell){$H$};
		
		
		\node (debut4) at (220pt, 4\wirespace){};
		\node (debut5) at (220pt, 3\wirespace){};
		\node [left = of debut4] (ket4) {$\ket 0$};
		\node [left = of debut5] (ket5) {$\ket 0$};
		\node[below right = -6pt and -36pt of debut4] (w4) {\scriptsize{4}};
		\node[below right = -6pt and -36pt of debut5] (w5) {\scriptsize{5}};
		
		\node[right=30pt of mes2, inner sep = 0] (coinx) {};
		\node[right=60pt of mes1, inner sep = 0] (coinz) {};
		
		\node[draw, circle, fill=black, minimum width=0.14cm, inner sep=0pt, right= 25pt + \messpace of Hbell] (C3) {}; 
		\draw let  
    			\p1 = (C3) in
      			(\x1,4\wirespace) circle (0.33\boxsize)
			node[inner sep = 0] (T3){}; 
		\draw let 
			\p1 = (T3) in
			(C3) -- (\x1, \y1+0.33\boxsize); 
			
		\node[draw, circle, fill=black, minimum width=0.14cm, inner sep=0pt, right= 55pt+\boxsize+2pt of T2] (C4) {}; 
		\draw let  
    			\p1 = (C4) in
      			(\x1, 3\wirespace) circle (0.33\boxsize)
			node[inner sep = 0] (T4){}; 
		\draw let 
			\p1 = (T4) in
			(C4) -- (\x1, \y1+0.33\boxsize); 
		
		\node[draw, rectangle, minimum width=\boxsize, minimum height=\boxsize, inner sep = 0, right=291pt of ket3] (X){$X$};
		\node[draw, circle, fill=black, minimum width=0.14cm, inner sep=0pt]  (C5) at ([yshift=3\wirespace]X){};
		\draw (C5) -- (X); 
			
		\node[draw, rectangle, minimum width=\boxsize, minimum height=\boxsize, right=321pt of ket3] (Z){$Z$};
		\node[draw, circle, fill=black, minimum width=0.14cm, inner sep=0pt]  (C6) at ([yshift=4\wirespace]Z){};
		\draw (C6) -- (Z); 
		
		\node[right = 20pt of Z] (fin3) {$\ket \psi$};
		\node (fin2) at ([yshift=\wirespace, xshift=-88pt]fin3) {\hphantom{$\ket \psi$}};
		\node (fin1) at ([yshift=\wirespace]fin2) {\hphantom{$\ket \psi$}};
		\node (fin5) at ([yshift=3\wirespace]fin3) {\hphantom{$\ket \psi$}};
		\node (fin4) at ([yshift=\wirespace]fin5) {\hphantom{$\ket \psi$}};

		\draw (ket1) -- (U);
		\draw (U) -- (Hbell);
		\draw (Hbell) -- (fin1);
		\draw (ket2) -- (H1);
		\draw (H1) -- (fin2);
		\draw (ket3) -- (X);
		\draw (X) -- (Z);
		\draw (Z) -- (fin3);
		\draw (ket4) -- (fin4);
		\draw (ket5) -- (fin5);
		
		  \node[inner sep = 0] at (40pt, 2.5\wirespace) (p1){};
		  \node[inner sep = 0] at (90pt, 0.4\wirespace) (p2){};
		  \node[inner sep = 0] at (253pt, 0.4 \wirespace)(p3){};
		  \node[inner sep = 0] at (253pt, 4.58\wirespace) (p4){};
		  
		  \node[below left = 0pt and 6pt of p4, inner sep = 0] (A) {$A$};
		  \node[below right = 0pt and 6pt of p4, inner sep = 0] (B) {$B$};

		 \draw[dashed] (p1) -- (p2); 
		 \draw[dashed] (p2) -- (p3); 
		 \draw[dashed] (p3) -- (p4); 
		 

		\node (marker3) at ([xshift=25pt, yshift=15pt]U){}; 
		\node (t3) at ([xshift=25pt, yshift=-2\wirespace-10pt]U) {\scriptsize{$t=3$}};
		\draw[dotted] (t3) -- (marker3);
		
		\node (marker5) at ([xshift=25pt, yshift=15pt]Hbell){}; 
		\node (t5) at ([xshift=25pt, yshift=-2\wirespace-10pt]Hbell) {\scriptsize{$t=5$}};
		\draw[dotted] (t5) -- (marker5);
		
		\node (marker7) at ([xshift=10pt, yshift=15pt]C3){}; 
		\node (t7) at ([xshift=10pt, yshift=-2\wirespace-10pt]C3) {\scriptsize{$t=7$}};
		\draw[dotted] (t7) -- (marker7);
		
		\node (marker9) at ([xshift=99pt, yshift=15pt]C3){}; 
		\node (t9) at ([xshift=99pt, yshift=-2\wirespace-10pt]C3) {\scriptsize{$t=9$}};
		\draw[dotted] (t9) -- (marker9);
		
	\end{tikzpicture}
	\caption{Teleportation in unitary quantum theory.}
\end{figure}
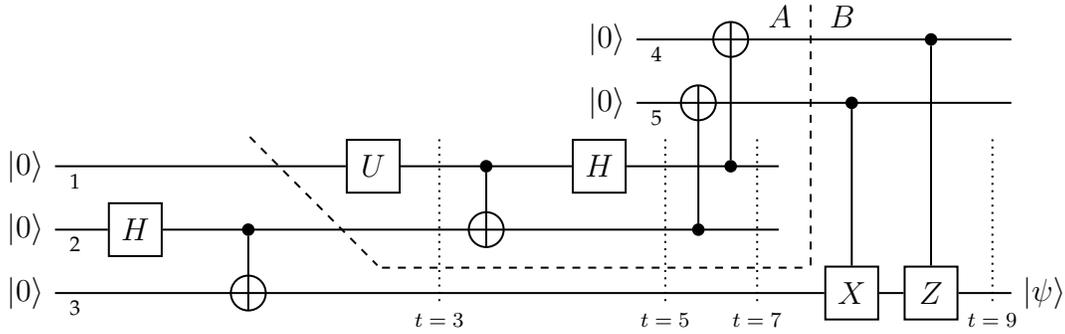
\label{fig:unitary}
On the question of the apparent instantaneous transfer in teleportation, what inhibits progress is the Schrödinger picture, for full unitarity does not 
change the fact that in $\ket{\Psi_3}$, $\alpha$ and $\beta$ can freely jump around the state vector. 
Therefore, on Schrödinger's stage, the local transfer remains unseeable, and one might be fooled by the suggestion that $\alpha$ and $\beta$ really are teleported. Heisenberg takes us backstage. 

\section{Heisenberg-Picture Descriptors}\label{sec:heis}

In this section, I present an overview of how descriptors work. For a thorough exposition, see~\cite{bedard2021abc}. 
A reader who prefers avoiding this section --- perhaps due to its more elaborate mathematical content --- can have a glance at equation~\eqref{eq:initialdescriptor}, accept the action of gates on descriptors as given by equations~\eqref{eqH},~\eqref{eqCnot} and~\eqref{eqU}, and proceed to the following section.

In the Heisenberg picture, observables evolve in time while the state remains constant; and conveniently set to $\ket{0}^{\otimes n}$ in the context of quantum computational networks. 
In front of the uncountable number of observables that evolve in time, the tempting thought is to give up on Heisenberg-picture descriptions. 
Fortunately, the algebra of observables admits a \emph{generating set}, i.e.~a set of operators that multiplicatively generate a basis of all operators.
This generating set can be chosen such that only a few operators act non-trivially on each system.
These operators are then encompassed into a single object, the \emph{descriptor} of the system, which evolves in time and can be used to calculate any time-evolved observable of that system. 
Similarly, any time-evolved observable that pertains to a collection of systems can be obtained from the descriptors of those systems.
To connect with the more familiar language of the Schrödinger picture, the descriptors corresponding to a collection of systems permit the reconstruction of the density matrix of this collection of systems.
In particular, the global density matrix can be obtained from all individual descriptors.
 
\subsection*{Qubit Descriptors}

Consider a network of~$n$ qubits, whose $i$\textsuperscript{th} qubit is denoted $\mathfrak Q_i$. 
At time~$0$ the descriptor of~$\mathfrak Q_i$ can be expressed as the pair of operators acting on~$(\mathbb C^2)^{\otimes n}$ 
\be \label{eq:initialdescriptor}
\bol q_i(0) = \left (q_{ix}(0),q_{iz}(0) \right )= 
 \left( \one^{i-1} \otimes \sigma_x \otimes \one^{n-i} 
\,,\, \one^{i-1} \otimes \sigma_z \otimes \one^{n-i} \right) \,,
\ee 
where~$\sigma_x$ and~$\sigma_z$ are Pauli matrices, and $\one^{k}$ is the identity on $(\mathbb C^2)^{\otimes k}$.
A third descriptor component,~$q_{iy}(0)$, can be obtained as~$iq_{ix}(0)q_{iz}(0)$.
Descriptors evolve as observables do; namely, if~$U$ denotes the evolution operator of what happens to the whole network between time~$0$ and time~$t$, then
\bea \label{eq:evotot} 
\bol q_i(t) = U^{\dagger} \bol q_i(0) U \,, 
\eea
where the~$U$ acts on both components of~$\bol q_i(0)$.
Time evolution preserves the algebra of descriptors, which, in the context of qubits, is the Pauli algebra,
\beas \label{eq:su2n}
[q_{iw} (t), q_{jw'}(t)] &=& 0\hphantom{q_{z}(t)} \qquad (i \neq j \text{ and } \forall w, w') \nonumber \\
q_{ix}(t) q_{iy}(t) &=& i q_{iz}(t)  \qquad(\text{and cyclic permutations})  \\
q_{iw}(t)^2 &=& \one\hphantom{q_{z}(t)} \qquad( \forall w)\,. \nonumber
\eeas

\sloppy

Suppose that  between the discrete times $t-1$ and $t$, only one gate is performed, whose matrix representation on the whole network is denoted~$G_t$. Therefore, $U= G_t V$, where $V$ consists of all gates from time $0$ to $t-1$. The evolution of descriptors can also be expressed in a step-by-step fashion, 
\be \label{eq:step}
\bol q_i(t) = \Uf^\dagger_{G_t}(\bol q(t-1)) \bol q_i(t-1) \Uf_{G_t}(\bol q(t-1)) \,,
\ee
where~$\bol q(\cdot)=(\bol q_1(\cdot), \dots, \bol q_n(\cdot))$ is the~$2n$-component object that encodes the descriptor of each qubit at the corresponding time and $\Uf_{G_t}(\cdot)$ is a fixed operator-valued function of some components of its argument. The function satisfies the defining equation $\Uf_{G_t}(\bol q(0))=G_t$, which is guaranteed to exist by the generative ability of the components of $\bol q(0)$.
More precisely, any linear operator~$G_t$ can be expressed as a polynomial in the~$2n$ matrices~$q_{1x}(0)$, \mbox{$q_{1z}(0)$, $\dots$,} $q_{nx}(0)$, $q_{nz}(0)$, and~$\Uf_{G_t}(\bol q(0))$ is one such polynomial.
The expressions \eqref{eq:evotot} and \eqref{eq:step} for the evolution of $\bol q_i(t)$ can be recognized equivalent: 
\beas
 V^\dagger G_t^\dagger  \bol q_i(0) G_t V 
&=& V^\dagger \Uf^\dagger_{G_t}(\bol q(0)) V V^\dagger \bol q_i(0)V V^\dagger \Uf_{G_t}( \bol q(0) ) V \\
&=& \Uf^\dagger_{G_t}\left (V^\dagger \bol q(0) V \right ) V^\dagger \bol q_i(0) V \, \Uf_{G_t}\left(V^\dagger \bol q(0) V \right)\\
&=& \Uf^\dagger_{G_t}\left(\bol q(t-1) \right) \bol q_i(t-1) \Uf_{G_t}(\bol q(t-1)) \,.
\eeas
The second equality follows because in each term of the polynomial $\Uf^\dagger_{G_t}\left (V^\dagger \bol q(0) V \right )$, products of components that are surrounded by $V^\dagger$ and $V$ will have their inner $V^\dagger$s and $V$s cancelled, leaving only the outer ones, which can be factorized outside of the polynomial to retrieve the first line. 

\subsection*{Locality and Completeness}

The \emph{locality} of the descriptors is due to the fact that if the gate~$G_t$ acts only on qubits of the subset $I\subset \{1,2,\ldots,n\}$, then its functional representation~$\Uf_{G_t}$ shall only depend on components of~$\bol q_k(t-1)$, for~$k \in I$. Therefore, for any~$j \notin I$, the 
descriptor~$\bol q_{j}(t-1)$ commutes with~$\Uf_{G_t}(\bol q(t-1))$, and thus remains unchanged between times~$t-1$ and~$t$: 
the description of system~$A$ is independent of what is done with system~$B$, which is dynamically isolated from the former.
Einstein's criterion, generalized as above with dynamical isolation instead of spatial separation, stresses that the locality of the mode of description is inherited from the locality of interactions.
If, in spacetime, the interactions are constrained by a lightcone structure, the descriptors inherit the constraint, and spatially separated systems shall be described independently of what is done to the other.

When the constant reference vector~$\ket{0}^{\otimes n} \equiv \, \ket{\bol 0}$ 
 is also taken into account, the descriptors are \emph{complete}. The expectation value~$\bra{\bol 0} \calO(t) \ket{\bol 0}$ of any observable~$\calO(t)$ that concerns only qubits of~$I$ can be determined by the descriptors~$\{\bol q_k(t)\}_{k\in I}$.
This can be seen more clearly at time~$0$, where an observable on the qubits of a subset~$I\subset \{1,2,\ldots,n\}$ is, like a gate~$G_t$, a linear 
operator that acts non-trivially \emph{only} on the qubits of~$I$. Again, any such operator can be generated additively and multiplicatively by the components of~$\bol q_k(0)$, with~$k\in I$. So there exists a polynomial $f_{\calO}$ such that~$\calO(0) = f_{\calO}(\{\bol q_k(0)\}_{k\in I})$. Therefore,
\bes
\calO(t) = U^\dagger \calO(0) U = f_{\calO}(\{U^\dagger \bol q_k(0) U\}_{k\in I}) = f_{\calO}(\{\bol q_k(t)\}_{k\in I})\,.
\ees
(Again, in a term which consists of a product of various components, the inner $U^\dagger$'s and $U$'s cancel out, and the outer $U^\dagger$'s and $U$'s can be factored out of the polynomial.) 
Since the elements of the density matrix can be computed as the expectation value of an appropriate operator, the density matrix~$\rho_I(t)$ of the joint subsystems in $I$ can be obtained from~$\{\bol q_k(t)\}_{k\in I}$.
In what follows, we shall only be interested in computing the reduced density matrix of one qubit from its descriptor.
The reduced density matrix~$\rho_k(t)= \tr{\comp{\mathfrak Q}_k}(U \ketbra{\bol 0}{\bol 0} U^\dagger)$ of~qubit $\mathfrak Q_k$ at time~$t$ can be expressed in the Pauli basis, like any $2 \times 2$ hermitian matrix of trace one, 
as
$$
\rho_k(t) = \frac 12 \left (\one + \sum_{w \in \{x,y,z\}} p_w(t)  \sigma_w \right ) \,.
$$
From the trace relations of Pauli matrices, the components~$p_{w}(t)$ are
\bes
p_{kw}(t) = \tr{}(\rho_k(t) \sigma_w) = \tr{}\left (U \ketbra{\bol 0}{\bol 0} U^\dagger (\one^{k-1} \otimes \sigma_w \otimes \one^{n-k}) \right) = \bra{\bol 0} q_{kw}(t)\ket{\bol 0} \,.
\ees 
The second equality comes from that $\rho^A \mapsto \rho^A \otimes \one^B$ is, as a super-operator, the adjoint of $\rho^{AB} \mapsto \tr{B} (\rho^{AB})$, and the rightmost equality follows from the cyclicity of the trace. That $q_{ky}(t)$ is not tracked in time is no problem since it can be computed as $iq_{kx}(t)q_{kz}(t)$.

\subsection*{The Action of Gates}

For concrete calculations in a network that admits a fixed set of gates, it is convenient to find the action that each gate has on the descriptors.
The matrix representation~$H_i$ of the Hadamard gate applied to~$\mathfrak{Q}_i$ (and the identity elsewhere) can be expressed as
\bes
H_i= \frac{q_{ix}(0) + q_{iz}(0)}{\sqrt 2} \,, 
\ees
which defines its functional representation~$\Uf_{H_i}(\bol q (\cdot))$. From equation~\eqref{eq:step}, and using the algebra of operators at time $t-1$ to simplify the right-hand side, one finds $\bol q_i(t) = \left(q_{iz}(t-1), q_{ix}(t-1)\right)$, or more elaborately,
$$
\left(q_{ix}(t-1), q_{iz}(t-1)\right) \equiv \bol q_i(t-1) \stackrel{H_i}{\to} \bol q_i(t)
 = \left(q_{iz}(t-1), q_{ix}(t-1)\right)\,.
$$ 
Therefore, the Hadamard gate switches the components of the descriptor on which it acts (regardless of how these components are expressed in terms of Pauli operators at time $t-1$).
Abstracting away the time at which the gate 
occurs, the action of~$H_i$ is specified by
\be \label{eqH}
H_i  \, \colon \,  \left(q_{ix}, q_{iz}\right) \to \left(q_{iz}, q_{ix}\right) \,.
\ee

In a like manner, the action of the~$\cnot$ can be found to be
\be \label{eqCnot}
\cnot  \, \colon \, 
\left\{ 
		\hspace{-5pt}
		\begin{array}{lcl}
			(q_{cx}  \hspace{-6pt}&,&\hspace{-6pt} q_{cz} ) \vspace{2pt}\\
			 (q_{tx}  \hspace{-6pt}&,&\hspace{-6pt} q_{tz} ) \vspace{2pt}\\
		\end{array}
		\hspace{-5pt}
	\right \} 
\to
\left \{ 
		\hspace{-5pt}
		\begin{array}{lcl}
			(q_{cx}  q_{tx}  \hspace{-6pt}&,&\hspace{-6pt} q_{cz}\hphantom{q_{tz}} ) \vspace{2pt}\\
			 (q_{tx}  \hspace{-6pt}&,&\hspace{-6pt} q_{cz} q_{tz} ) \vspace{2pt}\\
		\end{array}
		\hspace{-5pt}
	\right \} \,,
\ee
where the label $c$ refers to the $c$ontrol qubit and the label $t$, to the $t$arget qubit. The $z$-component of the control is copied onto the $z$-component of the target, while the $x$-component of the target is copied onto the $x$-component of the control (regardless of what those components are at that time).
The action of the controlled-$Z$ gate can be found from

$$
	
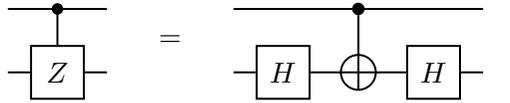
\begin{tikzpicture}  [scale=1, line width=0.75]
		\node[draw,rectangle,  minimum width=\boxsize, minimum height=\boxsize] 
		(Z) at (0,0){$Z$};
		\filldraw (0,\wirespace) circle (1.8pt);
		\draw (0,\wirespace) -- (Z.north);
		
		\node (bd) at (-0.8,0){};
		\node (bf) at (0.8,0){};
		\node (hd) at (-0.8,\wirespace){};
		\node (hf) at (0.8,\wirespace){};
		
		\draw (bd) -- (Z);
		\draw (Z) -- (bf);
		\draw (hd) -- (hf);
		
		\node at (1.5,0.5\wirespace){$=$};
		
		\node (bd2) at (2.2,0){};
		\node (bf2) at (5.8,0){};
		\node (hd2) at (2.2,\wirespace){};
		\node (hf2) at (5.8,\wirespace){};
		
		\node[draw, circle, fill=black, minimum width=0.14cm, inner sep=0pt] (C1) at (4,\wirespace){}; 
		\draw let  
    			\p1 = (C1) in
      			(\x1,0) circle (0.33\boxsize)
			node (T1){}; 
		\draw let 
			\p1 = (T1) in
			(C1) -- (\x1, \y1-0.33\boxsize); 

		\node[draw,rectangle, minimum width=\boxsize, minimum height=\boxsize] 
		(H1) at (3,0){$H$};
		
		\node[draw,rectangle, minimum width=\boxsize, minimum height=\boxsize] 
		(H2) at (5,0){$H$};
		
		\draw (bd2) -- (H1);
		\draw (H1) -- (H2);
		\draw (H2) -- (bf2);
		\draw (hd2) -- (hf2);
	\end{tikzpicture} \,.
$$

In the teleportation protocol, Alice's preparation consists of a generic one-qubit gate~$U$, which maps~$\ket 0$ to~\mbox{$\ket \psi = \alpha \ket 0 + \beta \ket 1$}. Such a generic transformation of $\text{SU}(2)$ can be expressed as the exponentiation of a generator,
$
U = e^{- i \frac{\phi}{2} \hat \phi \cdot \vec \sigma} \,,
$
or alternatively, 
it can be parametrized with Euler angles as
\be \label{eq:paramU}
U = 
e^{i \varphi_3 \sigma_z} e^{i \varphi_2 \sigma_x} e^{i \varphi_1 \sigma_z} 
= \begin{pmatrix}
e^{i (\varphi_1 + \varphi_3)} \cos \varphi_2 & i e^{-i(\varphi_1- \varphi_3)}\sin \varphi_2 \\
i e^{i(\varphi_1- \varphi_3)}\sin \varphi_2 & e^{-i (\varphi_1 + \varphi_3)} \cos \varphi_2
\end{pmatrix}\,.
\ee
Note that $\alpha$ and~$\beta$ are parametrized, as in the first column of~$U$, with respect to~\mbox{$\vec \varphi = (\varphi_1,\varphi_2,\varphi_3)$}.
Since Alice's preparation involves an action only on $\mathfrak Q_1$, it can be expressed as in equation~\eqref{eq:paramU} where the components of $\bol q_1(0)$ replace $\sigma_x$ and $\sigma_z$. This thus defines the functional representation of $\Uf_U(\bol q (\cdot))$ from which the action can be computed, 
\be \label{eqU}
U  \, \colon \,  \bol q_1 = \left(q_{1x}, q_{1z}\right) \to (\Dxq, \Dzq) \,.
\ee
Sparing the detailed expressions, $\Dxq$ and $\Dzq$ denote two functions which depend on the operators $q_{1x}$, $q_{1z}$ and on the parameters $\vec \varphi$ (and therefore on $\alpha$ and $\beta$). 

%
%

\section{The Heisenberg Picture of Teleportation}\label{sec:heistele}

Teleportation is now revisited in the Heisenberg picture.
Descriptors are \emph{such} a local description of quantum systems, that a computation in a network can be carried out by writing the descriptors directly on each qubit wire, as in~Figure~\ref{teleportation2}.

\begin{sidewaysfigure}
	\centering
	\begin{tikzpicture} [scale=1, line width=0.75]

		\node (debut3) at (0,0){};
		\node (debut2) at (0, \wirespaceh){};
		\node (debut1) at (0, 2\wirespaceh){};
		
		\node [left = of debut3] (ket3) {};
		\node [left = of debut2] (ket2) {};
		\node [left = of debut1](ket1) {};

		\node[below right = -6pt and -36pt of debut1] (w1) {\tiny{1}};
		\node[below right = -6pt and -36pt of debut2] (w2) {\tiny{2}};
		\node[below right = -6pt and -36pt of debut3] (w3) {\tiny{3}};

		\node[above right = -5pt and -33pt of debut1] (q10) {\scriptsize{$(q_{1x}, q_{1z})$}};
		\node[above right = -5pt and -33pt of debut2] (q20) {\scriptsize{$(q_{2x}, q_{2z})$}};
		\node[above right = -5pt and -33pt of debut3] (q30) {\scriptsize{$(q_{3x}, q_{3z})$}};
		
		\node[draw, rectangle, minimum width=\boxsize, minimum height=\boxsize, right=46pt of ket2] (H1){$H$};

		\node[right = 18pt of q20] (q21) {\scriptsize{$(q_{2z}, q_{2x})$}};
		
		\node[draw, circle, fill=black, minimum width=0.14cm, inner sep=0pt, right= 40pt of H1] (C1) {}; 
		\draw let  
    			\p1 = (C1) in
      			(\x1,0) circle (0.33\boxsize)
			node (T1){}; 
		\draw let 
			\p1 = (T1) in
			(C1) -- (\x1, \y1-0.33\boxsize); 
		
		\node[right = 3pt of q21] (q22) {\scriptsize{$(q_{2z}q_{3x}, q_{2x})$}};
		
		\node[right = 72pt of q30] (q32) {\scriptsize{$(q_{3x}, q_{2x}q_{3z})$}};
		
		\node[draw,rectangle, minimum width=\boxsize, minimum height=\boxsize, right=34pt of debut1] (U) {$U$};
		
		\node[right = 45pt of q10] (q11) {\scriptsize{$(\Dxq, \Dzq)$}};
		
		\node (marker3) at ([xshift=38pt, yshift=-30pt]U){}; 
		\node (t3) at ([xshift=0pt, yshift=-57pt]marker3) {\scriptsize{$t=3$}};
		\draw[dotted] (t3) -- (marker3);

		\node[draw, circle, fill=black, minimum width=0.14cm, inner sep=0pt, right= 75pt of U] (C2) {}; 
		\draw let  
    			\p1 = (C2) in
      			(\x1,\wirespaceh) circle (0.33\boxsize)
			node[inner sep = 0] (T2){}; 
		\draw let 
			\p1 = (T2) in
			(C2) -- (\x1, \y1-0.33\boxsize); 
		
		\node[right = 2pt of q11] (q14) {\scriptsize{$(\Dxq q_{2z}q_{3x}, \Dzq)$}};
		
		\node[right = 14pt of q22] (q25) {\scriptsize{$(q_{2z}q_{3x}, \Dzq q_{2x})$}};
	  
		\node[draw,rectangle, minimum width=\boxsize, minimum height=\boxsize, right = 95pt of C2] (Hbell){$H$};
		 
		\node[right = 18pt of q14] (q15) {\scriptsize{$(\Dzq, \Dxq q_{2z}q_{3x})$}};
		
		\node (debut4) at (250pt, 4\wirespaceh){};
		\node (debut5) at (250pt, 3\wirespaceh){};
		
		\node [left = of debut4] (ket4) {};
		\node [left = of debut5] (ket5) {};
		
		\node[below right = -6pt and -36pt of debut4] (w4) {\tiny{4}};
		\node[below right = -6pt and -36pt of debut5] (w5) {\tiny{5}};
		
		\node[above right = -5pt and -33pt of debut4] (q45) {\scriptsize{$(q_{4x}, q_{4z})$}};
		\node[above right = -5pt and -33pt of debut5] (q55) {\scriptsize{$(q_{5x}, q_{5z})$}};
		
		\node[draw, circle, fill=black, minimum width=0.14cm, inner sep=0pt, right= 345pt  of debut2] (C3) {}; 
		\draw let  
    			\p1 = (C3) in
      			(\x1,3\wirespaceh) circle (0.33\boxsize)
			node[inner sep = 0] (T3){}; 
		\draw let 
			\p1 = (T3) in
			(C3) -- (\x1, \y1+0.33\boxsize); 
			
		\node[draw, circle, fill=black, minimum width=0.14cm, inner sep=0pt, right= 345pt + \messpace of debut1] (C4) {}; 
		\draw let  
    			\p1 = (C4) in
      			(\x1, 4\wirespaceh) circle (0.33\boxsize)
			node[inner sep = 0] (T4){}; 
		\draw let 
			\p1 = (T4) in
			(C4) -- (\x1, \y1+0.33\boxsize); 
			
		\node (marker7) at ([xshift=288pt, yshift=0pt]marker3){}; 
		\node (t7) at ([xshift=0pt, yshift=-57pt]marker7) {\scriptsize{$t=7$}};
		\draw[dotted] (t7) -- (marker7);
			
		\node[right = 122pt of q25] (q27) {\scriptsize{$(q_{2z}q_{3x}q_{5x}, \Dzq q_{2x})$}};
		
		\node[right = 12pt of q15] (q17) {\scriptsize{$(\Dzq q_{4x}, \Dxq q_{2z}q_{3x})$}};
		
		\node[right = 105pt of q45] (q47) {\scriptsize{$(q_{4x}, \Dxq q_{2z}q_{3x}q_{4z})$}};
		
		\node[right = 99pt of q55] (q57) {\scriptsize{$(q_{5x}, \Dzq q_{2x}q_{5z})$}};
		
		\node[draw, rectangle, minimum width=\boxsize, minimum height=\boxsize, inner sep = 0, right=460pt of debut3] (X){$X$};
		\node[draw, circle, fill=black, minimum width=0.14cm, inner sep=0pt]  (C5) at ([yshift=3\wirespaceh]X){};
		\draw (C5) -- (X); 
				
		\node[draw, rectangle, minimum width=\boxsize, minimum height=\boxsize, right=20pt of X] (Z){$Z$};
		\node[draw, circle, fill=black, minimum width=0.14cm, inner sep=0pt]  (C6) at ([yshift=4\wirespaceh]Z){};
		\draw (C6) -- (Z); 
		
		\node (marker9) at ([xshift=447pt, yshift=0pt]marker3){}; 
		\node (t9) at ([xshift=0pt, yshift=-57pt]marker9) {\scriptsize{$t=9$}};
		\draw[dotted] (t9) -- (marker9);
		
		\node[below right = 7pt and 250pt of q32] (q38) {\scriptsize{$(q_{3x}, \Dzq q_{3z}q_{5z})$}};
		
		\node[below right = -5pt and -83pt of q38] (q39) {\scriptsize{$( \Dxq q_{2z}q_{4z}, \Dzq q_{3z}q_{5z})$}};

		\node[right = 460pt of debut1] (fin1) {};
		\node[right = 460pt of debut2] (fin2) {};
		\node[right = 535pt of debut3] (fin3) {};
		\node[right = 285pt of debut4] (fin4) {};
		\node[right = 285pt of debut5] (fin5) {};
		
		\node[inner sep = 0] (debutfleche) at ($(X)!0.5!(Z)$) {};
		\draw[>={Latex[length=1mm]}, <->, line width=0.1mm] (debutfleche)  to [out=260, in=0] (q38.east);
		
		\node[inner sep = 0] (debutfleche) at ($(Z)!0.50!(fin3)$) {};
		\draw[>={Latex[length=1mm]}, <->, line width=0.1mm] (debutfleche)  to [out=260, in=10] (q39.east);
		
		\draw (ket1) -- (U);
		\draw[mygreen, line width=0.6mm] (U) -- (C2);
		\draw[mygreen, line width=0.60mm] (C2) -- (Hbell);
		\draw[mygreen, line width=0.60mm] (Hbell) -- (C4);
		\draw[mygreen, line width=0.60mm] (C4) -- (fin1);

		\draw (ket2) -- (H1);
		\node[inner sep = 0] (endT2) at ([xshift=0.33\boxsize]T2){}; 
		\draw (H1) -- (endT2.east);
		\draw[mygreen, line width=0.60mm] (endT2) -- (C3);
		\draw[mygreen, line width=0.60mm] (C3) -- (fin2);
		
		\draw (ket3) -- (X);
		\draw[mygreen, line width=0.60mm] (X) -- (Z);
		\draw[mygreen, line width=0.60mm] (Z) -- (fin3);
		
		\node[inner sep = 0] (endT4) at ([xshift=0.33\boxsize]T4){}; 
		\draw (ket4) -- (endT4.east);
		\draw[mygreen, line width=0.60mm] (endT4) -- (C6);
		\draw[mygreen, line width=0.60mm] (C6) -- (fin4);
		
		\node[inner sep = 0] (endT3) at ([xshift=0.33\boxsize]T3){}; 
		\draw (ket5) -- (endT3.east);
		\draw[mygreen, line width=0.60mm] (endT3) -- (C5);
		\draw[mygreen, line width=0.60mm] (C5) -- (fin5);
		
		\draw let 
			\p1 = (T4) in
			(C4) -- (\x1, \y1+0.33\boxsize); 
		\draw (C6) -- (Z); 
	\end{tikzpicture}
	\vspace{10pt}
	\caption{The descriptors in quantum teleportation. The computation starts with all qubits properly initialized to~$
\bol q_i(0)$, which, for conciseness is denoted without time dependence as~$(q_{ix}, q_{iz})$.
When entering a gate, the components of the input are shuffled into the output in accordance with the action of the gate, which is prescribed by equations~\eqref{eqH},~\eqref{eqCnot} and~\eqref{eqU}. 
The information flow of the parameters~$\alpha$ and~$\beta$, encoded in $\vec \varphi$, is highlighted in green, with the wires thickened. 
It spreads locally in the network through the interactions, and, as can be seen, \emph{the ``classical'' bits are responsible for carrying the parameters encoding Alice's system over to Bob's location}.}
	\label{teleportation2}
\end{sidewaysfigure}
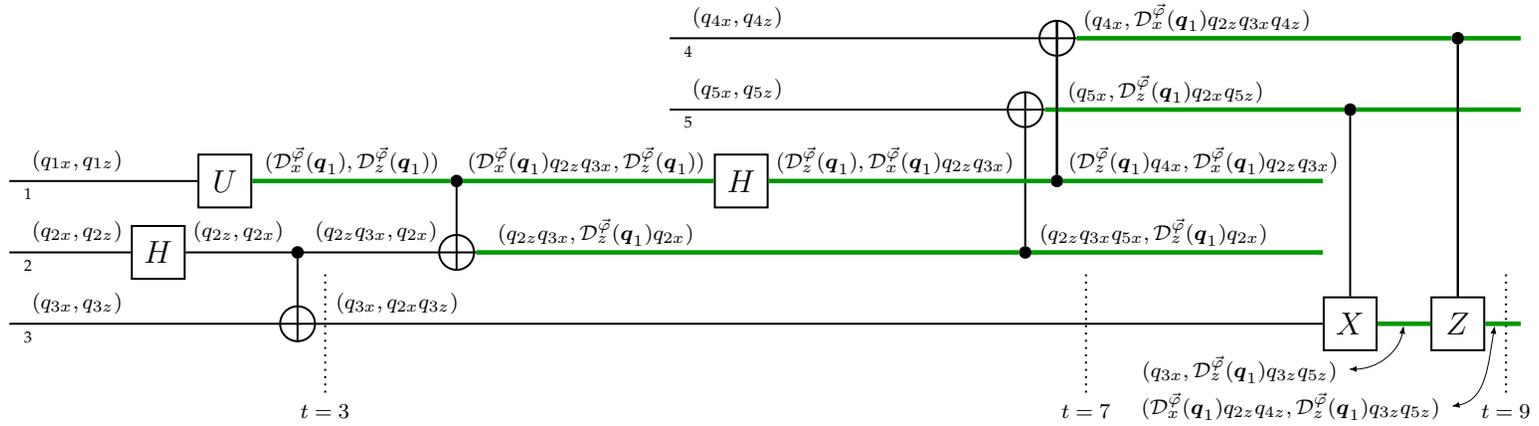


The correspondence with the famous result in the Schrödinger picture can be verified by computing the reduced density matrix of~$\mathfrak Q_3$ at time~$9$,~$\rho_3(9)$.
It is expressed by
$$
\rho_3(9) = \frac{\one}{2} + \frac{1}{2}\sum_{w \in \{x,y,z\}} \langle \bol 0| q_{3w}(9) |    \bol 0\rangle \ \sigma_w \,,
$$
where $ |  \bol 0\rangle \equiv \ket 0^{\otimes 5}$ the fixed Heisenberg state.
Because 
$$\bol q_{3}(9) = 	(\Dxq q_{2z}q_{4z} , \Dzq q_{3z}q_{5z})  
~~~\text{and}~~~ \bol q_{1}(3) = (\Dxq, \Dzq)$$ 
only differ by operators that have eigenvalues $1$ with respect to $\ket{ \bol 0 }$,
\be \label{eq:ev}
\langle   \bol 0| q_{3w}(9) |   \bol 0\rangle = \langle   \bol 0| q_{1w}(3) |   \bol 0\rangle \,.
\ee
Therefore,
$
\rho_3(9) = \rho_1(3) =\, \ket\psi \bra\psi \,;
$
the state vector that corresponds to Bob's descriptor after his correction is $\ket{\psi}$.

\subsection*{Locally Inaccessible Information}

To articulate the localization of information, Deutsch and Hayden did not require a quantitative notion of information. 
Instead, they coined and worked with the following criteria:
\begin{itemize}
\item[(i)] A system~$\mathfrak S$ is deemed to contain information about a parameter $\theta$ if (though not necessarily only if) the probability of some outcome of some measurement on~$\mathfrak S$ alone depends on~$\theta$;
\item[(ii)] A system~$\mathfrak S$ is deemed to contain no information about~$\theta$ if there exists a complete description of $\mathfrak S$ that satisfies Einstein’s criterion and is independent of~$\theta$.
\end{itemize}
Following these criteria, Deutsch and Hayden realized that there is such a thing as
\emph{locally inaccessible} information, namely, information which is present in a system but does not affect the probability of any outcome of any possible measurement on that system alone. 
Notably, in the communication channel used in teleportation,~$\alpha$ and~$\beta$ are locally inaccessible.
Indeed, the collection of systems~$\mathfrak Q_3$,~$\mathfrak Q_4$ and~$\mathfrak Q_5$ at time $7$ contains information about~$\alpha$ and~$\beta$ since, as the rest of the protocol shows, the parameters can crop up in the probability distributions of some measurement brought about only by those systems. 
However, by (ii),~$\alpha$ and~$\beta$ do not reside in~$\mathfrak Q_3$, for $\bol q_3(7)$ is independent of them. 
Therefore,~$\alpha$ and~$\beta$ are located in~$\mathfrak Q_4$ and~$\mathfrak Q_5$, notwithstanding their associated density matrix proportional to the identity.
Being locally inaccessible, the information about $\alpha$ and $\beta$ that is carried in~$\mathfrak Q_4$ and~$\mathfrak Q_5$ remains unaffected by measurements, and thus it remains unaffected by decoherence. 
This observation has prompted Deutsch and Hayden to realize the tradeoff between local accessibility and robustness to decoherence in 
information transfer.

\subsection*{On the Classicality of the Bits}
The discoverers of teleportation pointed out in the first sentence of their abstract that the process relies on ``purely classical information''. 
However, as a potential critique might have it, the use of the quantum bits~$\mathfrak Q_4$ and~$\mathfrak Q_5$ as a communication channel contradicts the requirement that classical bits are to be utilized.
Not only does this, \emph{prima facie}, undermines the very purpose of teleportation by seemingly having a quantum channel already in place, but it appears to be a flagrant category mistake: classical and quantum bits are of a fundamentally different kind, one might argue.

The claim of ``purely classical information'' is the crucial element in the conjuring trick's setup. 
Let us not be fooled by it, for
\emph{there is no such thing, fundamentally, as purely classical information}: either quantum theory holds universally, or it does not, but in the latter case, an explanatory theory about a boundary of its domain of validity is required. 
Everett's proposal~\cite{dewitt2015many} was that although measurement interactions seem to impose a boundary on the domain of unitary quantum theory, they do not. 
The key to unravelling teleportation is to accept that, by the same token, classical information also does not push against the domain of unitary quantum theory; rather, it is absorbed by it.
For a unified theory, the primary concern
is to explain ``purely classical information'' in terms of quantum systems~\cite{zurek2022quantum, wallace2012emergent} and not vice versa. 
Whatever it is that we view as purely classical information is instantiated in physical systems, which, after all, satisfy quantum theory.

Yet, it can be argued that, indeed, the two bits are classical, according to explanations of what ``classical'' can mean \emph{within} quantum theory.
First, as illustrated in Figure~\ref{phone}, a nearby environment can be modelled to decohere the two qubits in the basis that has been selected by the measurement interaction. 
The environment contains at least the logical space of two qubits, whose descriptors~$\bol q_E$ and~$\bol q_{E'}$ are given by some generic representation of the Pauli algebra, i.e. they need not be initialized as in equation~\eqref{eq:initialdescriptor}. 
When the environment is affected by the records $\mathfrak Q_4$ and $\mathfrak Q_5$,  the records are also affected: the $x$-components of the environment reach the $x$-components of the records.
However, these operators do not make it any further towards~$\mathfrak Q_3$, because the interactions that follow involve the records as control qubits, and so only pass on their $z$-components.
Therefore, a decohering environment does not prevent teleportation; the transfer is \emph{robust to decoherence}, a distinguishing property of classical communication.

Second, I suggested that the classical communication channel could be thought to be a telephone, but in the protocol displayed in Figure~\ref{teleportation2} the records $\mathfrak Q_4$ and $\mathfrak Q_5$ are physically brought to Bob's location.
 In usual classical communication channels, precise (quantum) systems are seldom sent from one location to another;
rather, the information is transmitted through a \emph{chain reaction} that occurs in a collection of quantum systems. 
 Therefore, a telephone, or more generally, the relevant degrees of freedom in classical communication, can be modelled by a chain reaction that involves the prepared systems $\mathfrak Q_{4'}$, $\mathfrak Q_{5'}$, $\mathfrak Q_{4''}$, $\mathfrak Q_{5''}$, and generically many others. 
See Figure~\ref{phone} for the calculation, which yield a final descriptor with dependencies on $q_{4'z}, q_{4''z}, q_{5'z}, q_{5''z}$. 
If the telephone line is properly initialized with descriptors of the 
form~\eqref{eq:initialdescriptor}, the operators appended to the final descriptor do not affect the expectation values in~\eqref{eq:ev} because they have eigenvalue 1 with respect to $\ket{\bol 0}$;
the Schrödinger state corresponding to $\mathfrak Q_3$ after the process remains~$\ket \psi$.
\emph{Thus, not only is the information transfer robust to decoherence, but it is realizable in a chain reaction that resembles a classical communication.}

Note, moreover, that decoherence can occur everywhere in the telephone line without inhibiting the transfer of $\Dxq$ and $\Dzq$ on Bob's final descriptor. 
As a final remark: where is Bob?
Isn't he supposed to receive the communication and act on $\mathfrak Q_3$ in accordance with the received bits?
Yes, but this is taken into account in the generically many other systems involved in the communication line.
No special assumptions are required to include Bob in Figure \ref{phone}: only his ability to manipulate systems mechanically and decoherently, like any other parts of the communication line. 

\begin{figure}
\begin{tikzpicture} [scale=1,  line width=0.75] 
		\centering
			
		\node (debut3) at (0,0){};
		\node (debut5) at (0, 3\wirespacehb + 2 \wirespaceht){};
		\node (debut4) at ([yshift=\wirespaceht]debut5){};
			
		\node[above right = -5pt and -33pt of debut3] (q35) {\scriptsize{$(q_{3x}, q_{2x}q_{3z})$}};
		
		\node[above right = -5pt and -33pt of debut4] (q45) {\scriptsize{$(q_{4x}, \Dxq q_{2z}q_{3x}q_{4z})$}};
		\node[above right = -5pt and -33pt of debut5] (q55) {\scriptsize{$(q_{5x}, \Dzq q_{2x}q_{5z})$}};

		\node[draw, circle, fill=black, minimum width=0.14cm, inner sep=0pt, right= 60pt of debut5] (CE') {}; 
		\draw let  
    			\p1 = (CE') in
      			(\x1, \y1+\wirespacehb+\wirespaceht) circle (0.33\boxsize)
			node[inner sep = 0] (TE'){}; 
		\draw let 
			\p1 = (TE') in
			(CE') -- (\x1, \y1+0.33\boxsize); 
		
		\node[draw, circle, fill=black, minimum width=0.14cm, inner sep=0pt, right= 60pt+\messpace of debut4 ] (CE) {}; 
		\draw let  
    			\p1 = (CE) in
      			(\x1,\y1+\wirespacehb+\wirespaceht) circle (0.33\boxsize)
			node[inner sep = 0] (TE){}; 
		\draw let 
			\p1 = (TE) in
			(CE) -- (\x1, \y1+0.33\boxsize); 
			
		\node (env') at ([xshift=-30pt]TE'){};
		\node (env) at ([yshift=\wirespaceht]env'){};
		
			\node[above right = -5pt and -33pt of env] (qE) {\scriptsize{$(q_{Ex}, q_{Ez})$}};
		\node[above right = -5pt and -33pt of env'] (qE') {\scriptsize{$(q_{E'x}, q_{E'z})$}};

		\node[draw, circle, fill=black, minimum width=0.14cm, inner sep=0pt, right= 25pt of CE] (C1) {}; 
		\draw let  
    			\p1 = (C1) in
      			(\x1,\y1-\wirespacehb-\wirespaceht) circle (0.33\boxsize)
			node[inner sep = 0] (T1){}; 
		\draw let 
			\p1 = (T1) in
			(C1) -- (\x1, \y1-0.33\boxsize); 
			
		\node[draw, circle, fill=black, minimum width=0.14cm, inner sep=0pt, right= 25pt of CE'] (C2) {}; 
		\draw let  
    			\p1 = (C2) in
      			(\x1,\y1-\wirespacehb-\wirespaceht) circle (0.33\boxsize)
			node[inner sep = 0] (T2){}; 
		\draw let 
			\p1 = (T2) in
			(C2) -- (\x1, \y1-0.33\boxsize); 
			
		\node[above right = -2pt and 18pt of C1] (q47) {\scriptsize{$(q_{4x}q_{Ex}, \Dxq q_{2z}q_{3x}q_{4z})$}};
		\node[above right = -2pt and 18pt of C2] (q57) {\scriptsize{$(q_{5x}q_{E'x}, \Dzq q_{2x}q_{5z})$}};
		
		\node[inner sep = 0] (debutfleche) at ($(CE)!0.5!(C1)$) {};
		\draw[>={Latex[length=1mm]}, <->, line width=0.1mm] (debutfleche)  to [out=40, in=180] (q47.west);
		
		\node[inner sep = 0] (debutfleche) at ($(CE')!0.5!(C2)$) {};
		\draw[>={Latex[length=1mm]}, <->, line width=0.1mm] (debutfleche)  to [out=40, in=180] (q57.west);

		\node (debut4') at ([xshift=-40pt]T1){};
		\node (debut5') at ([yshift=-\wirespaceht]debut4'){};

		\node[below right = -6pt and -36pt of debut4'] (w4') {\tiny{4'}};
		\node[below right = -6pt and -36pt of debut5'] (w5') {\tiny{5'}};
		
		\node[above right = -5pt and -33pt of debut4'] (q45') {\scriptsize{$(q_{4'x}, q_{4'z})$}};
		\node[above right = -5pt and -33pt of debut5'] (q55') {\scriptsize{$(q_{5'x}, q_{5'z})$}};
		
		\node[draw, circle, fill=black, minimum width=0.14cm, inner sep=0pt, right= 25pt of T1] (C3) {}; 
		\draw let  
    			\p1 = (C3) in
      			(\x1,\y1-\wirespacehb-\wirespaceht) circle (0.33\boxsize)
			node[inner sep = 0] (T3){}; 
		\draw let 
			\p1 = (T3) in
			(C3) -- (\x1, \y1-0.33\boxsize); 
			
		\node[draw, circle, fill=black, minimum width=0.14cm, inner sep=0pt, right= 25pt of T2] (C4) {}; 
		\draw let  
    			\p1 = (C4) in
      			(\x1,\y1-\wirespacehb-\wirespaceht) circle (0.33\boxsize)
			node[inner sep = 0] (T4){}; 
		\draw let 
			\p1 = (T4) in
			(C4) -- (\x1, \y1-0.33\boxsize); 
			
		\node[above right = -2pt and 18pt of C3] (q46') {\scriptsize{$(q_{4'x}, \Dxq q_{2z}q_{3x}q_{4z}q_{4'z})$}};
		\node[above right = -2pt and 18pt of C4] (q56') {\scriptsize{$(q_{5'x}, \Dzq q_{2x}q_{5z}q_{5'z})$}};
		
		\node[inner sep = 0] (debutfleche) at ($(T1)!0.5!(C3)$) {};
		\draw[>={Latex[length=1mm]}, <->, line width=0.1mm] (debutfleche)  to [out=40, in=180] (q46'.west);
		
		\node[inner sep = 0] (debutfleche) at ($(T2)!0.5!(C4)$) {};
		\draw[>={Latex[length=1mm]}, <->, line width=0.1mm] (debutfleche)  to [out=40, in=180] (q56'.west);
		
		\node (debut4'') at ([xshift=-45pt]T3){};
		\node (debut5'') at ([yshift=-\wirespaceht]debut4''){};

		\node[below right = -6pt and -36pt of debut4''] (w4'') {\tiny{4''}};
		\node[below right = -6pt and -36pt of debut5''] (w5'') {\tiny{5''}};

		\node[above right = -5pt and -33pt of debut4''] (q45'') {\scriptsize{$(q_{4''x}, q_{4''z})$}};
		\node[above right = -5pt and -33pt of debut5''] (q55'') {\scriptsize{$(q_{5''x}, q_{5''z})$}};
		
		\node[right = 25pt of T4, draw, circle, fill=black, minimum width=0.14cm, inner sep=0pt]  (C5) {};
		\node[draw, rectangle, minimum width=\boxsize, minimum height=\boxsize, inner sep = 0] (X) at ([yshift=-\wirespacehb]C5){$X$};
		\draw (C5) -- (X); 
		
		
		\node[draw, rectangle, minimum width=\boxsize, minimum height=\boxsize, inner sep = 0, right=5pt of X] (Z){$Z$};
		\node[draw, circle, fill=black, minimum width=0.14cm, inner sep=0pt]  (C6) at ([yshift=\wirespacehb+\wirespaceht]Z){};
		\draw (C6) -- (Z); 
		
		\node[above right = -2pt and 18pt of C6] (q4'') {\scriptsize{$(q_{4''x}, \Dxq q_{2z}q_{3x}q_{4z}q_{4'z}q_{4''z})$}};
		\node[above right = -2pt and 33pt of C5] (q5'') {\scriptsize{$(q_{5''x}, \Dzq q_{2x}q_{5z}q_{5'z}q_{5''z})$}};
		
		\node[inner sep = 0] (debutfleche) at ($(T3)!0.5!(C6)$) {};
		\draw[>={Latex[length=1mm]}, <->, line width=0.1mm] (debutfleche)  to [out=40, in=180] (q4''.west);
		
		\node[inner sep = 0] (debutfleche) at ($(T4)!0.5!(C5)$) {};
		\draw[>={Latex[length=1mm]}, <->, line width=0.1mm] (debutfleche)  to [out=40, in=180] (q5''.west);

		\node[right = 370pt of debut3] (fin3) {};
		
		\node[right = 235pt of debut4] (fin4) {};
		\node[right = 235pt of debut5] (fin5) {};
		
		\node[right = 235pt of debut4'] (fin4') {};
		\node[right = 235pt of debut5'] (fin5') {};

		\node[right = 235pt of debut4''] (fin4'') {};
		\node[right = 235pt of debut5''] (fin5'') {};
		
		\node[right = 180pt of env] (finE) {};
		\node[right = 180pt of env'] (finE') {};
		
		\node[right = 165pt of q35] (q392) {\scriptsize{$(\Dxq q_{2z}q_{4z}q_{4'z}q_{4''z},\Dzq q_{3z}q_{5z}q_{5'z}q_{5''z})$}};
		
		\node[below right = -6pt and -36pt of debut3] (w3) {\tiny{3}};
		
		\node[below right = -6pt and -36pt of debut4] (w4) {\tiny{4}};
		\node[below right = -6pt and -36pt of debut5] (w5) {\tiny{5}};
		
		\node[below right = -6pt and -36pt of env] (wE) {\tiny{E}};
		\node[below right = -6pt and -36pt of env'] (wE') {\tiny{E'}};		
		
		\node [left = of debut3] (ket3) {};
		\node[left = of env] (kete) {};
		\node[left = of env'] (kete') {};
		\node [left = of debut4] (ket4) {};
		\node [left = of debut5] (ket5) {};
		\node[left = of debut4'] (ket4') {};
		\node[left = of debut5'] (ket5') {};
		\node[left = of debut4''] (ket4'') {};
		\node[left = of debut5''] (ket5'') {};
		
		\draw (ket3) -- (X);
		\draw (X) -- (Z);
		\draw (Z) -- (fin3);
		
		\draw (ket4) -- (fin4);
		\draw (ket5) -- (fin5);
		
		\draw (ket4') -- (fin4');
		\draw (ket5') -- (fin5');
		
		\draw (ket4'') -- (fin4'');
		\draw (ket5'') -- (fin5'');
		
		\draw (kete) -- (finE);
		\draw (kete') -- (finE');

	\end{tikzpicture}
	\caption{A telephone.}
	\label{phone}
\end{figure}
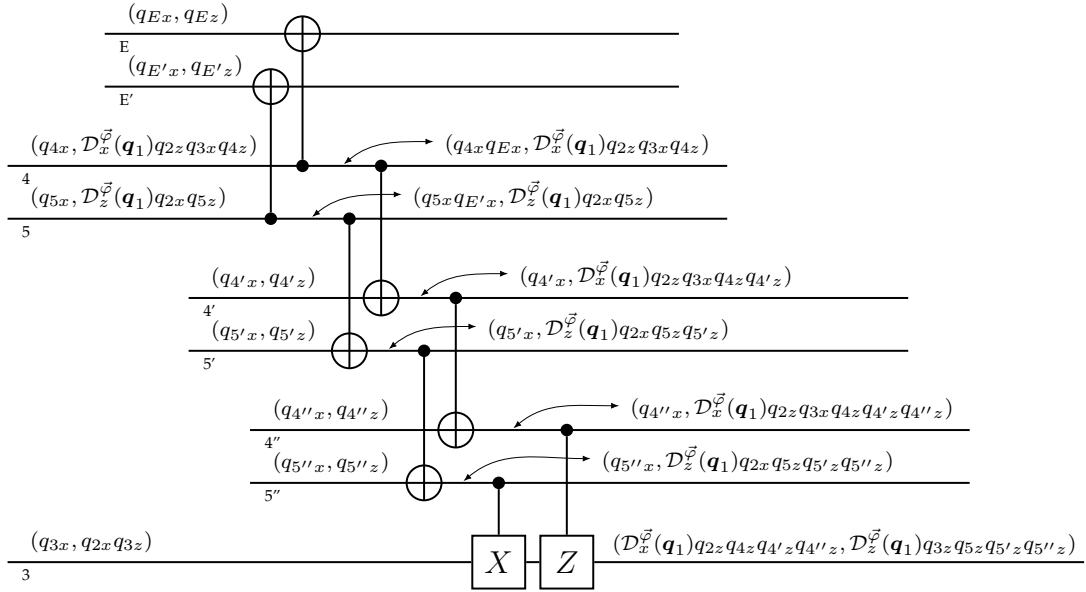

\subsection*{Explaining the Information Transfer}

So, how do $\alpha$ and $\beta$ make their way from Alice's location to Bob's?
The Copenhagen tradition would have it due to the collapse of the state vector, which has prompted many to view the transfer as instantaneous, through action at a distance. 
Jozsa suggested that ‘‘nonlocal influences’’ allow them to ‘‘fly across the entanglement’’~\cite{jozsa1998pages}.
%
For Penrose, the entangled pair has a channel which ‘‘proceeds into the past’’~\cite{penrose1998quantum} and into the future again.
In the Bohmian theory, the ‘‘transfer is mediated by the nonlocal quantum potential’’~\cite{maroney1999quantum}.
Vaidman suggests that ``the nonlocality of Everett’s world\footnote{As Popescu mentionned~\cite{MWIday6}, these worlds should be called \emph{``Lev's worlds''}, for they extend arbitrarily far in space and are constructed from so-called ``macroscopic objects'' in a ``definite classically described state''~\cite{vaidman2021many}, but no 
 signs of such concepts at the fundamental level can be found in Everett's writings.} is the basis of the teleportation of quantum information''~\cite{vaidman2022transfer}. 
Most proponents of unitary quantum theory in the Schrödinger picture do not resign to such extravagant conjectures, for they do not attempt to amend quantum theory. 
Braunstein argued that ``the quantum information is `hidden' within the correlations between the system and the environment while being wholly absent from any of the individual subsystems''~\cite{braunstein1996quantum}, and 
Timpson wrote that 
 ``global rather than local properties are being used to carry information''~\cite{timpson2006grammar}.
%
These proposals are confined by the unfulfillment of Einstein's criterion of locality by the global state vector, which prevents the localization of the parameters in the Schrödinger picture.
But they can be localized in the Heisenberg picture.
According to Deutsch and Hayden, the information about~$\alpha$ and~$\beta$ is transported ``simply, prosaically, in the qubits $\mathfrak Q_4$ and $\mathfrak Q_5$ as they travelled from $A$ to $B$''.

\section{Counterfactual Elements of Reality}
\label{sec:morethan}

Despite being expressed in terms of two complex numbers, the state prepared at Alice's location, $\alpha \ket 0 + \beta \ket 1$, contains only two free (real) parameters due to the constraints imposed by normalization and the irrelevance of the phase factor.
The latter can be considered by demanding that $\alpha = |\alpha|$, which also fixes~$|\beta|$ through normalization, so viewed in this way, the second free parameter is the phase of~$\beta$, $\arg (\beta)$.
Alice's preparation~$U$, however, is a generic one-qubit gate, which, as parametrized by $\vec \varphi$ in equation~\eqref{eq:paramU}, contains three\footnote{
In fact, a generic element of $\textrm{U}(2)$ contains four free parameters. However, since $U \in \textrm{SU}(2)$, $\det U = 1$, which almost amounts to quotiating the global phase of the unitary operator. ``Almost'', because there remains a $\mathbb Z_2$ ambiguity due to a possible factor $- \one$, which leaves the determinant equal to $1$ while being a non-trivial global phase. However, the possibility for this factor can be avoided by suitable constraints on the domain of $\vec \varphi$.} free parameters.
Since $U$ can be expressed as
\beas \label{eq:paramU2}
U
&=&
e^{i (\varphi_1 + \varphi_3)}
\begin{pmatrix}
\cos \varphi_2 &  e^{-2i (\varphi_1+ \varphi_3)}i e^{2i \varphi_3} \sin \varphi_2 \\
i e^{-2i \varphi_3}\sin \varphi_2 & e^{-2i (\varphi_1 + \varphi_3)} \cos \varphi_2
\end{pmatrix}\\
&\equiv&
(\eta^*)^{1/2}
\begin{pmatrix}
\alpha &  -\eta \beta^* \\
\beta & \eta \alpha 
\end{pmatrix} \,,
\eeas
where $\alpha \in \mathbb R^+$ and $\beta$ are the amplitudes of the prepared state, and
the extra parameter~$\eta$, of unit norm, labels a one-parameter family of states that are legitimate images of~$\ket 1$ under~$U$: each of the $\eta (\beta^* \ket 0 + \alpha \ket 1)$ are orthogonal to $\alpha \ket 0 + \beta \ket 1$. 
Therefore, the descriptor's components~$\Dxq$ and~$\Dzq$ that depend on~$\vec \varphi$ 
can alternatively be thought to depend on~$|\alpha|$, $\arg (\beta)$ and $\eta$.
Since the descriptor of~$\mathfrak Q_3$ at time $9$ also carries the dependency on~$\eta$, a question arises: \emph{has $\eta$ also reached to Bob's location?} 

\subsection*{The Instrumentalist Temptation}
Deutsch and Hayden's criteria remain silent on the question of the localization of $\eta$.
Indeed, even if all systems are collected at some time after the preparation, there exists no measurement on the network as a whole whose distribution of outcomes would depend on $\eta$.
Therefore, $\eta$ fails to fulfill criterion (i), even if $\mathfrak S$ is taken to be the network as a whole.
A tempting view is to dismiss the existence of all that is oblivious from experiments, which embodies the instrumentalist attitude, namely, the consideration of scientific theories as mere tools for predictions.
In spite of Deutsch and Hayden's warning ``(though not necessarily only if)'', which insists that criterion (i) is only sufficient, 
instrumentalism would also demand it be
necessary in a strong sense,
namely, it might demand that if the distribution of some measurement outcomes on~$\mathfrak S$ alone is independent of~$\theta$, then~$\theta$ is not a descriptive element of~$\mathfrak S$. 
Not only would~$\eta$ be deemed to be absent from the system, but~$\alpha$ and~$\beta$, too, could not be thought to be localized in~$\mathfrak Q_4$ and~$\mathfrak Q_5$ as they are transferred.
The Heisenberg-picture description would vanish, for what remains after the instrumentalist's mutilation would be informationally equivalent to the global state, and 
Raymond-Robichaud~\cite[\S4]{raymond2021local} showed that any attempt to build a local and complete description of quantum systems from the state vector alone must fail\footnote{In Timpson's terminology~\cite{timpson2005nonlocality}, the instrumentalist's mutilation amounts to shifting from the ontological to the conservative interpretation of Heisenberg-picture descriptors. In Raymond-Robichaud's~\cite{raymond2021local}, from noumenal to phenomenal states.}. 

In his thesis, Everett criticizes instrumentalism~\cite{dewitt2015many}:
 \begin{quote}
It is necessary to say a few words about a view which is sometimes expressed, the idea that a physical theory should contain no elements which do not directly correspond to observables. This position seems to be founded on the notion that the only purpose of a theory is to serve as a summary of known data, and overlooks the second major purpose, the discovery of totally new phenomena. The major motivation for this viewpoint appears to be the desire to construct perfectly ``safe'' theories which will never be open to contradiction. Strict adherence to such a philosophy would probably seriously stifle the progress of physics.
\end{quote}

To embrace the full power of Heisenberg-picture descriptors is not merely to view them as another way to think of quantum theory, which may be convenient in some cases --- for instance, to make sense of teleportation. 
Nor is it 
a tool whose sole purpose is to make predictions.
It is to consider them as an account of reality.
The reality captured by the descriptors is larger than that captured by the universal state vector~\cite{bedard2021cost, 
wallacetimpson07, deutsch2011vindication}. 
In particular, it has room for $\eta$. 
The descriptive elements which, like $\eta$, lie in Heisenberg-picture descriptions but not in the Schrödinger state are globally inaccessible (not just locally).
They reside in the multiverse yet, in some unobservable sector, for only the sector which is singled out by the Heisenberg state is amenable to observations. 
The unobservable sector encompasses $\eta$ and, more generally, all that resides in the global unitary operator that embodies the dynamics and yet is beyond the ``column'' selected by the Heisenberg state.
All of it is \emph{counterfactual descriptive elements of reality}; it accounts for what would be accessible had some prior operation been performed.
In the teleportation setting,~$\eta$
 would have cropped up in the distributions had $\mathfrak Q_1$ been rotated anyhow except around the $Z$ axis before being prepared with~$U$.

\section{Conclusion}\label{sec:conclusion}

The solution by Deutsch and Hayden to the problem of teleportation provides a probe into the classical realm, which signals that it is much deeper than expected. 
In fact, it is even deeper than expected from the Schrödinger picture of unitary quantum theory.
Anyone who takes for granted that communication between Alice and Bob involves ``purely classical information'' is fooled by teleportation. 
The classical realm is quantum; 
a classical communication channel is one that is robust to decoherence and realizable in a chain reaction in quantum systems. 

Explaining classical communication from some interaction within quantum systems might seem radical at first glance.
But the opposite is true. 
If one posits that quantum theory does not universally hold, then one must explain where its boundary resides and why. 
The proposal here simply follows Everett's program to take the quantum theory seriously and, in the absence of the need to introduce a boundary to its domain of applicability, consider it universal. 

Unitarity does not fully clarify the explanation of teleportation in the Schr\"odinger picture. 
The explanation presented here is only possible in the Heisenberg picture of unitary quantum theory. 
Those accustomed to unitary quantum theory (i.e. Everettian quantum theory) shall see arguments for adopting and further developing the Heisenberg picture. 
But those who are still agnostic about how to ``interpret'' quantum theory --- namely, still deciding whether unitary quantum theory needs to be truncated, merged with another theory, or completed in some way --- will see in the proposed explanation of teleportation arguments for both the Heisenberg picture \emph{and} for unitary quantum theory, as their conjunction solves the problem of the locality of information transfer in teleportation.
Progress can be assessed by the problems that are solved.
When I explained the teleportation protocol in the Heisenberg picture to Gilles Brassard, one of the discoverers of the phenomenon, he right away told me that it was the most satisfactory elucidation of teleportation that he had ever heard. 
I hope that this piece can have a similar effect on you.

Teleportation is not the only phenomenon whose apparent non-locality has been puzzling.
Heisenberg-picture descriptors also make locality manifest in superdense coding~\cite{bedard2021abc} and in the Aharanov-Bohm effect~\cite{marletto2020aharonov}.

I shall conclude with a brief reflection on Lev's problem, \emph{why is Everettian quantum theory not in the consensus?}
Deutsch wrote~\cite{deutsch2011beginning}
\begin{quote}
Some people may enjoy conjuring tricks without ever wanting to know how they work. Similarly, during the twentieth century, most philosophers, and many scientists, took the view that science is incapable of discovering anything about reality. Starting from empiricism, they drew the inevitable conclusion (which would nevertheless have horrified the early empiricists) that science cannot validly do more than predict the outcomes of observations, and that it should never purport to describe the reality that brings those outcomes about. This is known as instrumentalism. 
\end{quote}

The prevalence of instrumentalism might be a part of the explanation as to why Everettian quantum theory is not in the consensus.
The denial of taking a theory seriously as a tentative account of reality also denies a proper investigation of its consequences. 
And satisfaction with mere predictions entails satisfaction with conjuring tricks: 
why should we strive for an explanation of teleportation when we already have a theory that predicts Bob's observations after the protocol?

\section{Discussion}\label{sec:discussion}

\nib{Lev Vaidman: }If I understand correctly, the story, your story, is about the universe. When we talk about teleportation, we talk about our world. And in the many world's interpretation, there is a part that concerns the whole universe. It is the part of the MWI where there is no collapse. There is no collapse here, no question. But there are no many worlds. Many worlds is when I perform my measurement, I split the world.
In teleportation, in every world, $\alpha$ and $\beta$ jump on Bob's qubit, and they jump at the moment of the measurement. So there is no other explanation within the world.

\medskip
\nib{CAB: }In the teleportation protocol, the records of the measurement eventually affect --- and get entangled with --- many other record-like systems, as well as many systems in the environment. 
In the Schrodinger picture, this leads to a wave function with four highly entangled terms, which, for all practical purposes, can no longer interact with one another via quantum interference: each term becomes autonomous.
What is more, in each term, there are relative properties between systems, which give a consistent account of what resembles a quasi-classical single ``world''.
This is the quantum theory of Everett, the unitarily evolving universal wave function, with important analyses further developed by Zeh, Zurek, Gell-Mann, Hartle, Saunders, Wallace and others.

Is it also yours?
It appears to me that you grant fundamental importance to notions like ``macroscopic objects'' in a ``definite classically described state'' from which you define your ``worlds'' (as the totality of all such objects)~\cite{vaidman2021many}.
The importance that you grant those words is also manifest in your attempt at explaining teleportation within those worlds 
as if the universal wave function was not a fundamental description but just a convenient way to stack all those worlds together.
Moreover, defining worlds via intuitive appeals to classicality and macroscopicity leads you close to a collapse theory; 
%
and in both cases, one is forced to suggest that in teleportation,~$\alpha$ and~$\beta$ jump to Bob's qubit.

To come back to descriptors, they also admit a decomposition into a sum of \emph{relative descriptors}, which, like in the Schrödinger picture, account for relative properties between systems. Yet, descriptors are foliated locally. See~\cite{kuypers2021everettian}.

\medskip
\nib{Andrew Jordan:} Let me make a critical comment. 
You made the claim that the bits that are transmitted between Alice and Bob by the telephone are really secretly quantum bits, and I must object to that because I think that if you claim that, then the logical consequence is that there is no such thing as classical information theory.
I think you have to give up on classical information theory as a thing that exists.
You say that, really, everything is quantum information theory.
But there are classical bits. 
And we are communicating with a classical channel, and so there are classical channels. 
So how do you respond to that criticism?

\newpage 
\nib{CAB:} Classical information theory can still exist; however, not fundamentally.
In fact, we know well how it is instantiated as a subcase of quantum information theory: a decohered state has a diagonal reduced density matrix whose numbers form a distribution, so we speak of its Shannon information.
However, this misses my point.
I took it as a premise that the world is quantum. 
A telephone is made of quantum systems. 
And yes, it looks like it is classical to me, but that is the program launched by Everett, namely, to understand how the quantum theory can explain the emergence of the classical.

\medskip
\nib{Simon Saunders: }Rather similar question:
I do not quite get it. 
The classical channel is really just telling Bob what the outcome of Alice's Bell measurement is. 
There is very little information there, whereas $\alpha$ and $\beta$ encode potentially vast amounts of information.
So I just don't quite get that. 
Can you elaborate?

\medskip
\nib{CAB: }The channel is indeed telling Bob what the outcome of Alice's Bell measurement is.
But it is not \emph{just} doing that.
Wouldn't you grant that any sort of classical channel that we can imagine is ultimately made of quantum systems?
\emph{This is not an irrelevant fact when we are trying to solve the capacity problem of teleportation.}
The quantum systems involved in the communication line transfer $\alpha$ and $\beta$ in a way that is locally inaccessible, resilient to decoherence and realizable in a chain reaction.

\medskip
\nib{Eric Curiel: }I have a quite general question about the approach.
How should I understand entanglement entropy in this picture?
It plays a very fundamental role from condensed matter physics to black hole thermodynamics.
How should I understand what seems to be the manifest non-locality of quantum mechanics that makes the efficacy of Von Neumann entanglement entropy possible?

\medskip
\nib{CAB: }Since density matrices can be recovered from descriptors and the constant Heisenberg state, so can Von Neumann entropy. 
But one way to understand entanglement between systems that is more in line with the Heisenberg picture is that no observable of a subsystem has a definite outcome, while some observables on the joint system do.
For instance, the preparation of~$\ket{\Phi^+}$ on~$\mathfrak Q_2$ and~$\mathfrak Q_3$ yields the following descriptors (see Figure \ref{teleportation2}):
$$ \bol q_{2}(3) = (q_{2z}(0)q_{3x}(0), q_{2x}(0)) \qquad \text{and} \qquad \bol q_{3}(3) = (q_{3x}(0), q_{2x}(0)q_{3z}(0)) \,.
$$
None of the observables corresponding to the descriptor components have a definite outcome since their expectation value 
 is $0$ while their spectrum is~$\pm 1$. This also holds for $q_{2y}(3)$ and $q_{3y}(3)$, and so for any observable obtained as a linear expansion of the descriptor components. However, $q_{2x}(3)q_{3x}(3)$, $- q_{2y}(3)q_{3y}(3)$ and $q_{2z}(3)q_{3z}(3)$ have a definite outcome:~$1$.

\medskip
\nib{Tim Maudlin: }I have two comments of a different character. One is just coming back to this telephone. There is only one information theory; it is Shannon information. You can apply it to bits, which by definition have only two possible states, you can apply it to spin $1/2$ particles that have infinitely many possible states, given by your $\alpha$ and $\beta$. It doesn't change information theory at all. In this protocol, all that is required to implement the protocol are two bits. That is all that is required. You may say, ``Oh, but I have to send a quantum system physically because physics is quantum mechanics.'' It doesn't matter if Alice sends a note classically; of course, it has more than two states, right? She can write in cursive, or she can write this way, so what? The point is that the protocol merely demands that you resolve between four possibilities. That requires two bits of information. Period end of the story.

The other comment is: the reason they gave that Nobel prize was for tests of violation of Bell's inequality at spacelike separation. That's the reason they say it shows non-locality. Quantum teleportation is puzzling, but one thing it sure doesn't do is violate Bell's inequality at spacelike separation. So to say, even if it were true, ``I have a local understanding of teleportation'' would not at all have any influence on the reason they say those experiments are so important.

\medskip
\nib{CAB: }The two comments are not of a different character: 
they answer one another. 
The primality of Shannon's information in one's mind makes one uncritical not only of its use in teleportation but also of the way in which the assumptions are coined in Bell's theorem, namely, in terms of classical probability distributions.
For whom the very use of classical probability distributions is not considered to be an assumption made by Bell, then indeed, the violation of Bell's inequality at spacelike separation challenges locality. 
Otherwise, the violation simply dismisses the hypothesis that quantum theory can be underlain by classical probability distributions.
%

\medskip
\nib{David Wallace:} I want to go back to the Deutsch--Hayden claim that once I have a local formulation of the theory, then the theory is local. 
The worry is that there are relatively clear cheap ways of making a theory local.
I'm not claiming this is a cheap way. But there are cheap ways.
For instance, I can just attach a copy of the state of the universe to every local system, I can say whatever my wildly nonlocal description is, in my new theory, the state of the system is the ordered pair of the state of the old theory and the state of the universe.
It is horrendously expensive;
call that a monadology move, for Leibniz's fans.
That framework is formally going to be local, but clearly, it is not telling us that the theory is interestingly local.
I don't think that the framework of descriptors has this character, although there are bits of it that sometimes worry me.
But I just want to flag that a bit more needs to be done to clarify that a theory is local just because it has a local formalism.
I think we have to avoid making moves of that kind.

\medskip
\nib{CAB:} What you suggest does not fulfill Einstein's criterion because if a state of the whole universe is included in the description of each localized system, then if Bob performs an operation on his system, it affects Alice's description. 

\subsection*{Aknowledgements}

I am deeply grateful to
Gilles Brassard, 
Xavier Coiteux-Roy,
David Deutsch,
Samuel Kuypers,
Jordan Payette,
William Schober,
Stefan Wolf
and an anonymous referee
for stimulating discussions or comments on earlier versions of this article.
This work was supported by the Fonds de recherche du Québec – Nature et technologie and the Swiss National Science Foundation.

\bibliographystyle{unsrt}
\bibliography{/Applications/TeX/ref}

\begin{thebibliography}{10}

\bibitem{notloacllyreal}
Daniel Garisto.
\newblock The universe is not locally real, and the physics {N}obel {P}rize
  winners proved it.
\newblock {\em Scientific American}, 328(1), 2022.

\bibitem{bell1964}
John~S Bell.
\newblock {On the Einstein Podolsky Rosen paradox}.
\newblock {\em Physics}, 1(3):195--200, 1964.

\bibitem{schilppalbert1970}
Paul~A Schilpp.
\newblock {\em Albert {E}instein: philosopher-scientist}, volume~7.
\newblock Open Court Publishing Co., 3d revised edition, 1970.

\bibitem{deutsch2000information}
David Deutsch and Patrick Hayden.
\newblock Information flow in entangled quantum systems.
\newblock {\em Proceedings of the Royal Society A. Mathematical, Physical and
  Engineering Sciences}, 456(1999):1759--1774, 2000.

\bibitem{healey1991holism}
Richard~A Healey.
\newblock Holism and nonseparability.
\newblock {\em The Journal of Philosophy}, 88(8):393--421, 1991.

\bibitem{wallace2010quantum}
David Wallace and Christopher~G Timpson.
\newblock Quantum mechanics on spacetime i: Spacetime state realism.
\newblock {\em The British journal for the philosophy of science}, 2010.

\bibitem{raymond2021local}
Paul Raymond-Robichaud.
\newblock A local-realistic model for quantum theory.
\newblock {\em Proceedings of the Royal Society A}, 477(2250):20200897, 2021.

\bibitem{raymond2017equivalence}
Paul Raymond-Robichaud.
\newblock The equivalence of local-realistic and no-signalling theories.
\newblock {\em arXiv preprint arXiv:1710.01380}, 2017.

\bibitem{bedard2021cost}
Charles~Alexandre B{\'e}dard.
\newblock The cost of quantum locality.
\newblock {\em Proceedings of the Royal Society A}, 477(2246):20200602, 2021.

\bibitem{gottesman1998heisenberg}
Daniel Gottesman.
\newblock The {H}eisenberg representation of quantum computers.
\newblock {\em arXiv preprint quant-ph/9807006}, 1998.

\bibitem{heisenberg1925quantum}
Werner Heisenberg.
\newblock Quantum-theoretical re-interpretation of kinematic and mechanical
  relations.
\newblock {\em Z. Phys}, 33:879--893, 1925.

\bibitem{bennett1993teleporting}
Charles~H Bennett, Gilles Brassard, Claude Cr{\'e}peau, Richard Jozsa, Asher
  Peres, and William~K Wootters.
\newblock {Teleporting an unknown quantum state via dual classical and
  Einstein-Podolsky-Rosen channels}.
\newblock {\em Physical Review Letters}, 70(13):1895, 1993.

\bibitem{bedard2021abc}
Charles~Alexandre B{\'e}dard.
\newblock The {ABC} of {D}eutsch--{H}ayden descriptors.
\newblock {\em Quantum Reports}, 3(2):272--285, 2021.

\bibitem{dewitt2015many}
Bryce~S DeWitt and Neill Graham.
\newblock {\em The many worlds interpretation of quantum mechanics}.
\newblock Princeton University Press, 1973.

\bibitem{zurek2022quantum}
Wojciech~Hubert Zurek.
\newblock Quantum theory of the classical: Einselection, envariance, quantum
  {D}arwinism and extantons.
\newblock {\em Entropy}, 24(11):1520, 2022.

\bibitem{wallace2012emergent}
David Wallace.
\newblock {\em The Emergent Multiverse: Quantum Theory According to the
  {E}verett Interpretation}.
\newblock Oxford University Press, 2012.

\bibitem{jozsa1998pages}
R~Jozsa.
\newblock Pages 369-379 in the geometric universe edited by s. huggett, l.
  mason, kp tod, st tsou and n. woodhouse, 1998.

\bibitem{penrose1998quantum}
Roger Penrose.
\newblock Quantum computation, entanglement and state reduction.
\newblock {\em Philosophical Transactions of the Royal Society of London.
  Series A: Mathematical, Physical and Engineering Sciences},
  356(1743):1927--1939, 1998.

\bibitem{maroney1999quantum}
O~Maroney and Basil~J Hiley.
\newblock Quantum state teleportation understood through the bohm
  interpretation.
\newblock {\em Foundations of Physics}, 29(9):1403--1415, 1999.

\bibitem{MWIday6}
The many-worlds interpretation of quantum mechanics day 6, October 2022.

\bibitem{vaidman2021many}
Lev Vaidman.
\newblock Many-worlds interpretation of quantum mechanics.
\newblock {\em Zalta {EN}, editor}, 2021.

\bibitem{vaidman2022transfer}
Lev Vaidman.
\newblock Transfer of quantum information in teleportation.
\newblock 2022.

\bibitem{braunstein1996quantum}
Samuel~L Braunstein.
\newblock Quantum teleportation without irreversible detection.
\newblock {\em Physical Review A}, 53(3):1900, 1996.

\bibitem{timpson2006grammar}
Christopher~G Timpson.
\newblock The grammar of teleportation.
\newblock {\em The British Journal for the Philosophy of Science}, 2006.

\bibitem{timpson2005nonlocality}
Christopher~G Timpson.
\newblock Nonlocality and information flow: The approach of {D}eutsch and
  {H}ayden.
\newblock {\em Foundations of Physics}, 35(2):313--343, 2005.

\bibitem{wallacetimpson07}
David Wallace and Christopher~G Timpson.
\newblock Non-locality and gauge freedom in {D}eutsch and {H}ayden's
  formulation of quantum mechanics.
\newblock {\em Foundations of Physics}, 37(7):1069--1073, 2007.

\bibitem{deutsch2011vindication}
David Deutsch.
\newblock Vindication of quantum locality.
\newblock {\em Proceedings of the Royal Society A. Mathematical, Physical and
  Engineering Sciences}, 468(2138):531--544, 2011.

\bibitem{marletto2020aharonov}
Chiara Marletto and Vlatko Vedral.
\newblock Aharonov-{B}ohm phase is locally generated like all other quantum
  phases.
\newblock {\em Physical Review Letters}, 125(4):040401, 2020.

\bibitem{deutsch2011beginning}
David Deutsch.
\newblock {\em The beginning of infinity: Explanations that transform the
  world}.
\newblock Penguin UK, 2011.

\bibitem{kuypers2021everettian}
Samuel Kuypers and David Deutsch.
\newblock Everettian relative states in the {H}eisenberg picture.
\newblock {\em Proceedings of the Royal Society A}, 477(2246):20200783, 2021.

\end{thebibliography}

\end{document}